\documentclass[conference]{IEEEtran}
\IEEEoverridecommandlockouts
\usepackage{cite}
\usepackage{amsmath,amssymb,amsfonts}
\usepackage{graphicx}
\usepackage{textcomp}
\usepackage{xcolor}
\def\BibTeX{{\rm B\kern-.05em{\sc i\kern-.025em b}\kern-.08em
    T\kern-.1667em\lower.7ex\hbox{E}\kern-.125emX}}

\usepackage[hidelinks]{hyperref}

\newcommand{\linestack}[1]{\def\arraystretch{0.9}\begin{tabular}[c]{@{}c@{}} #1 \end{tabular}}

\usepackage[hang,small,bf]{caption}
\usepackage[labelformat=simple]{subcaption}

\usepackage[whole]{bxcjkjatype}

\usepackage{multirow}
\usepackage{physics}
\usepackage{bm}

\usepackage{quantikz}
\usepackage{tikz}

\usepackage{algorithm}
\usepackage{algorithmicx}
\usepackage{algpseudocode}
\algnewcommand{\Initialize}[1]{
  \State \textbf{Initialize:}
  \State \hspace*{\algorithmicindent}\parbox[t]{0.8\linewidth}{\raggedright #1}
}

\begin{document}

\title{Feed-Forward Probabilistic Error Cancellation with Noisy Recovery Gates\\
}

\author{\IEEEauthorblockN{Leo Kurosawa}
\IEEEauthorblockA{\textit{Graduate School of Computer Science and Engineering} \\
\textit{University of Aizu}\\
Aizu-Wakamatsu, Fukushima, Japan \\
m5271018@u-aizu.ac.jp}\\

\IEEEauthorblockN{Xinwei Lee}
\IEEEauthorblockA{\textit{School of Computing and Information Systems}\\
\textit{Singapore Management University}\\
80 Stamford Rd, Singapore \\
xwlee@smu.edu.sg}\\

\IEEEauthorblockN{Ningyi Xie}
\IEEEauthorblockA{\textit{Graduate School of Science and Technology} \\
\textit{University of Tsukuba}\\
Tsukuba, Ibaraki, Japan \\
nyxie@cavelab.cs.tsukuba.ac.jp}\\

\IEEEauthorblockN{Jungpil Shin}
\IEEEauthorblockA{\textit{School of Computer Science and Engineering} \\
\textit{University of Aizu}\\
Aizu-Wakamatsu, Fukushima, Japan \\
jpshin@u-aizu.ac.jp}\\

\and

\IEEEauthorblockN{Yoshiyuki Saito}
\IEEEauthorblockA{\textit{Graduate School of Computer Science and Engineering} \\
\textit{University of Aizu}\\
Aizu-Wakamatsu, Fukushima, Japan \\
d8241104@u-aizu.ac.jp}\\

\IEEEauthorblockN{Xinjian Yan}
\IEEEauthorblockA{\textit{Graduate School of Science and Technology} \\
\textit{University of Tsukuba}\\
Tsukuba, Ibaraki, Japan \\
yanxinjian@cavelab.cs.tsukuba.ac.jp}\\

\IEEEauthorblockN{Dongsheng Cai}
\IEEEauthorblockA{\textit{Faculty of Engineering, Information and Systems} \\
\textit{University of Tsukuba}\\
Tsukuba, Ibaraki, Japan \\
cai@cs.tsukuba.ac.jp}\\

\IEEEauthorblockN{Nobuyoshi Asai}
\IEEEauthorblockA{\textit{School of Computer Science and Engineering} \\
\textit{University of Aizu}\\
Aizu-Wakamatsu, Fukushima, Japan \\
nasai@u-aizu.ac.jp}\\
}

\maketitle

\begin{abstract}
Probabilistic Error Cancellation (PEC) aims to improve the accuracy of expectation values for observables.
This is accomplished using the probabilistic insertion of recovery gates, which correspond to the inverse of errors.
However, the inserted recovery gates also induce errors. 
Thus, it is difficult to obtain accurate expectation values with PEC since the estimator of PEC has a bias due to noise induced by recovery gates.
To address this challenge, we propose an improved version of PEC that considers the noise resulting from gate insertion, called Feed-Forward PEC (FFPEC). 
FFPEC provides an unbiased estimator of expectation values by cancelling out the noise induced by recovery gates.
We demonstrate that FFPEC yields more accurate expectation values compared to the conventional PEC method through analytical evaluations.
Numerical experiments are used to evaluate analytical results.
\end{abstract}


\section{Introduction}
Fascinated by the possibilities of high-speed computing with quantum computers, many quantum algorithms have been proposed to tackle problems that require a lot of calculation time even in classical computers, such as quantum chemical calculations \cite{peruzzo2014Variationala, fedorov2022VQE} and combinatorial optimization problems \cite{farhi2014Quantumb, bartschi2020Grover}.
These algorithms require expectation values of some observables to obtain a solution.
However, since a Noisy Intermediate-Scale Quantum (NISQ) computer \cite{preskill2018Quantum} is not equipped with error correction systems for noises, the accuracy of an expectation value obtained from measurement outcomes of a quantum computer is inevitably degraded by the noises.
To obtain a meaningful result from a NISQ computer, quantum error mitigation (QEM) techniques without additional qubits have been proposed to suppress the effects of noises \cite{temme2017Error, vandenberg2022Modelfreea, berg2023Probabilistic, endo2018Practical, hicks2022Active, maciejewski2020Mitigation, bravyi2021Mitigating, guo2022Quantum, gupta2023Probabilistic, kim2020Quantuma, koczor2024Probabilistic, majumdar2022Error, nation2021Scalable, nguyen2023Information}.

Probabilistic Error Cancellation (PEC) is one of the most well-known QEM techniques used to reduce noises caused by quantum gate operations \cite{temme2017Error, berg2023Probabilistic, guo2022Quantum, koczor2024Probabilistic, gupta2023Probabilistic}.
The key idea of PEC is to cancel out noise $\Lambda$ induced by a quantum gate operation by adding an operation corresponding to the inverse of the noise.
This operation to cancel the noise is represented as a linear combination of quantum gates.
To cancel out the noise in this way, PEC requires a full characterization of the noise such as its probability to occur and its inverse.
If we have the full knowledge of the noise, it is possible to obtain an unbiased estimator for the expectation value of an ideal quantum circuit, through statistical post-processing of measurement outcomes.

Using PEC, an unbiased estimator for an ideal expectation value can be obtained with a variance of $O(\gamma^2)$, where $\gamma \geq 1$ is a factor that increases exponentially with the number of quantum gates in a quantum circuit.
To obtain an expectation value with a high accuracy, a large number of shots is essential to reduce the variance.
So, this $\gamma$ is called the{\it sampling overhead}.
From a practical standpoint, it is important to reduce the value of $\gamma$.
Some methods have been proposed to minimize the sampling overhead $\gamma$ by decomposing a quantum circuit with a set of gate operations that a quantum computer supports \cite{piveteau2022Quasiprobability, takagi2021Optimal}.

The characterization of noise and the reduction of sampling overhead described above are necessary to obtain more accurate expectation values with the PEC.
However, in practice, there is a limit to the accuracy of expectation values that can be obtained even if the noise can be well-characterized because recovery gates (gates inserted to cancel noise) induce noise again.
Thus, the PEC estimator has a bias due to noise induced by applying the recovery gate \cite{cai2023Quantum, jin2024Noisy}.
Since recovery gates are also run on noisy quantum computers, it is difficult to reduce this bias to zero.
Exploring methods to reduce the bias of the PEC estimator to zero is necessary to obtain an accurate expectation value, while reluctantly accepting noises caused by recovery gates.

In this paper, we study a method to achieve zero bias in the PEC estimator even with noises induced by the recovery gate applied.
We propose an error mitigation method, named Feed-Forward PEC (FFPEC), which addresses noises induced by recovery gate operations in the PEC process to reduce the bias to zero.
The key idea of FFPEC is to cancel out noises induced by the recovery gates by defining an inverse map of noise including these noises.
Although the sampling cost $\gamma$ of FFPEC is slightly larger than that of the standard PEC, the expectation values for the FFPEC estimator are expected to be closer to an exact expectation value than for the standard PEC.
FFPEC can be viewed as a concrete implementation of noise cancellation maps for practical PEC, which is also considered in \cite{jin2024Noisy}.
We demonstrate the performance of FFPEC by comparing the standard PEC using analytical and numerical results with the depolarizing noise \cite{nielsen2010Quantuma}, which is studied in some error mitigation methods \cite{vovrosh2021Simple, urbanek2021Mitigating, temme2017Error, jin2024Noisy}.

The present paper is organized as follows. First, we give a background for the standard PEC technique and depolarizing noise model in Sec.~\ref{sec:preliminary}.
In Sec.~\ref{sec:FFPEC}, we present the Feed-Forward PEC.
Then, numerical experiments are conducted to evaluate the performance of FFPEC by comparing it with the standard PEC in Sec.~\ref{sec:numexp}.
Finally, we conclude about FFPEC in Sec.~\ref{sec:conculusions}.

\section{Background} \label{sec:preliminary}
We first briefly introduce the Probabilistic Error Cancellation (PEC) \cite{temme2017Error, suzuki2022Quantum}.
We also introduce a commonly used noise model, depolarizing, and its inverse for PEC.

\subsection{Probabilistic Error Cancellation}
We first introduce the notations required to explain the PEC procedure.
We consider an $n$ qubits system.
Let $\mathcal{Q}$ denote a unitary trace-preserving completely positive (TPCP) map of an ideal quantum circuit on the $n$ qubits system, $\rho_\text{in}$ denote the input quantum state to this circuit $\mathcal{Q}$ and $\rho_{\text{out}}$ denote the output quantum state after applying the circuit $\mathcal{Q}$ to the input state $\rho_\text{in}$ with a noisy quantum computer.
The circuit $\mathcal{Q}$ can be represented as $\mathcal{Q}(\rho) = Q\rho Q^\dagger$ with a corresponding unitary matrix $Q$ for any quantum state $\rho$.
In addition, the quantum circuit $\mathcal{Q}$ consists of $N_G$ ideal quantum gates as follows:
\begin{equation}
    \mathcal{Q} := \mathcal{U}_{N_G} \circ \mathcal{U}_{N_G-1} \circ \cdots  \mathcal{U}_{2} \circ  \mathcal{U}_{1},
\end{equation}
where $\circ$ represents a composition of TPCP maps, and each $\mathcal{U}_i$ is a unitary TPCP map corresponding to a quantum gate $U_i$.
Suppose that a quantum circuit $\mathcal{Q}$ is executed on a noisy quantum computer and that each noise channel $\Lambda_i$ is induced for each gate $\mathcal{U}_i$, where noise channel $\Lambda_i$ can be assumed as a Pauli channel \cite{wallman2016Noise, berg2023Probabilistic}.
The output quantum state $\rho_{\text{out}}$ is written as:
\begin{equation}
    \rho_{\text{out}} =  \Lambda_{N_G} \circ \mathcal{U}_{N_G} \circ \Lambda_{N_G -1} \circ \mathcal{U}_{N_G -1} \cdots \Lambda_{1} \circ \mathcal{U}_{1} (\rho_{\text{in}}).
\end{equation}
We denote an ideal output quantum state as $\rho_{\text{out}}^{\text{ideal}} := \mathcal{Q}(\rho_{\text{in}})$, then an ideal expectation value of an observable $M$ is written as $\ev{M}_{\text{ideal}} := \Tr[M \rho_{\text{out}}^{\text{ideal}}]$.

The objective of PEC is to provide an expectation value with good accuracy while executing a quantum circuit on a noisy quantum computer.
The key idea of PEC is to cancel out noises $\Lambda_i$ which occurs after $U_i$ is executed by adding its inverse $\Lambda_i^{-1}$.
By cancelling out noises in such a way, PEC can provide an expectation value $\ev{M}_\mathcal{Q}$ via statistical post-processing.
To simplify, we describe the PEC process using a single quantum gate $U$.
After that, the PEC process is explained for a general quantum circuit.
First of all, let $\Lambda$ be a noise induced by operating a quantum gate $\mathcal{U}$.
We assume that there exists a gate set $\{ \mathcal{R}_i \}$ that is implemented on a quantum computer so that the inverse map of noise $\Lambda^{-1}$ can be represented as a linear combination with elements of this set as a basis:
\begin{equation}
    \Lambda^{-1} = \sum_{i} q_i \mathcal{R}_i, \label{eq:Lambda_inverse}
\end{equation}
where each $q_i$ is a real number (pseudo-probability) and satisfies $\sum_i q_i = 1$ \cite{temme2017Error}.
Then we have
\begin{equation}
    \mathcal{U} = \Lambda^{-1} \Lambda \mathcal{U} = \sum_i q_i \mathcal{R}_i \Lambda \mathcal{U}
                = \gamma \sum_i \sigma_i {\rm sgn}(q_i) \mathcal{R}_i \Lambda \mathcal{U}, \label{eq:PEC_for_SingleGate}
\end{equation}
with $\gamma = \sum_i \abs{q_i}$ and $ \sigma_i = \abs{q_i} / \gamma$.
Factor $\gamma$ is the sampling overhead, and each $\sigma_i$ is the insertion probability of gate $\mathcal{R}_i$, and ${\rm sgn}(q_i) = \pm 1$ means a parity corresponding to gate $\mathcal{R}_i$.

Using $\gamma$, the expectation value of observable $M$ can be expressed as follows:
\begin{equation}
    \ev{M}_{\mathcal{U}} = \gamma \sum_i \sigma_i \ev{\mu^\text{eff}_i},
\end{equation}
where $\mu_i^{\text{eff}}={\rm sgn}(q_i)m_i$ and $m_i$ is the measurement outcome of the state $\mathcal{R}_i \Lambda \mathcal{U}(\rho_{\text{in}})$, which is generated by applying the gate $\mathcal{R}_i$ with probability $\sigma_i$ after applying gate $\mathcal{U}_i$.
Multiplying the outcome $\mu_i$ by the parity ${\rm sgn}(q_i)$ corresponding to $\mathcal{R}_i$ gives $\mu_i^\text{eff}$.
The expectation value of the random variable $\gamma\mu^{\text{eff}}$ is approximately an unbiased estimator of $\ev{M}_{\mathcal{U}}$.

To obtain an expectation value $\ev{M}_\mathcal{Q}$ of an observable $M$ with a quantum circuit $\mathcal{Q}$, we apply the process stated above for $\mathcal{Q}$.
Applying the above process to the quantum circuit $\mathcal{Q} = \prod_{k=1}^{N_G}\mathcal{U}_k$, we have
\begin{equation}
    \prod_{k=1}^{N_G} \mathcal{U}_k = \prod_{k=1}^{N_G} \gamma^{(k)} \sum_{i_1, \cdots, i_{N_G}} \prod_{k=1}^{N_G} {\rm sgn}(q_{i_{k}}) \prod_{k=1}^{N_G} \sigma_{i_{k}} \prod_{k=1}^{N_G} \mathcal{R}_{i_{k}} \Lambda_k \mathcal{U}_k,
    \label{eq:sum_Uk}
\end{equation}
with $\mathcal{U}_k = \gamma^{(k)}\sum_{i_k}\mathrm{sgn}(q_{i_k})\sigma_{i_k}\mathcal{R}_{i_k}\Lambda_k\mathcal{U}_k$.
We can obtain a random variable $\mu^{\text{eff}}$ by executing a circuit modified by operation $\mathcal{R}_{i_k}$ added with probability $\sigma_{i_k}$ and multiplying sign $\prod_{k=1}^{N_G} {\rm sgn}(q_{i_{k}})$ to measurement outcome.
The expectation value $\ev{\mu^{\text{eff}}}$ is obtained by repeating this process, and finally we obtain the expectation value $\ev{M}_\mathcal{Q} = \gamma^{\text{tot}}\ev{\mu^{\text{eff}}}$, where $\gamma^{\text{tot}} = \prod_{k=1}^{N_G} \gamma^{(k)}$.
The estimator $\ev{M}_{\mathcal{Q}}$ is unbiased for the exact expectation value $\ev{M}_{\text{ideal}}$.

Here, operation $\mathcal{R}$ added to cancel noise is called recovery gate in this paper.
Note that the recovery gate $\mathcal{R}$ can sometimes introduce noises.
To avoid this noise, the recovery $\mathcal{R}$ can be merged with the neighboring gate.
However, the gate merging would break the consistency with the PEC process in practical uses.
Thus, the gate merging is not performed in PEC.

\subsection{Depolarizing noise}

We introduce depolarizing \cite{nielsen2010Quantuma}, which is used in our experiments.
To explain the noise channel, we denote $\mathcal{I}, \mathcal{X}, \mathcal{Y}$, and $\mathcal{Z}$ as trace-preserving and completely positive (TPCP) maps corresponding to the Pauli matrices $I, X, Y$, and $Z$, respectively.
Similarly, $\mathcal{II}, \mathcal{IX}, \ldots, \mathcal{ZZ}$ also represent TPCP maps corresponding to couples of Pauli matrices $II, IX, \ldots, ZZ$.
In addition, a single qubit state and a two qubits state are denoted as $\rho$ and $\rho'$, respectively.
The error rate is denoted as $p$.

\subsubsection{Depolarizing for single qubit state}
Let $\mathcal{D}_1$ denote a TPCP map of the depolarizing noise on a single qubit state $\rho$ as follows:
\begin{equation}
    \mathcal{D}_1(\rho) = \left(1-\frac{3p}{4}\right)\mathcal{I}(\rho)  + \frac{p}{4} (\mathcal{X}(\rho) + \mathcal{Y}(\rho) + \mathcal{Z}(\rho)).
\end{equation}
The inverse map of $\mathcal{D}_1$ is written as follows \cite{guo2023Noise}:
\begin{equation}
    \mathcal{D}_1^{-1}(\rho) = \left(1-\frac{3q}{4}\right)\mathcal{I}(\rho)  + \frac{q}{4} (\mathcal{X}(\rho) + \mathcal{Y}(\rho) + \mathcal{Z}(\rho)),
\end{equation}
where
\begin{equation}
    q = \frac{-p}{1-p}.  \label{eq:pec:1depo:q}
\end{equation}
The sampling cost is $\gamma=(1+p/2)/(1-p)$, which gives that the insertion probabilities of Pauli gates (recovery gates) $X, Y$, and $Z$ are all the same value $p/(4+2p)$.

\subsubsection{Depolarizing for two qubits state}
Let $\mathcal{D}_2$ denote a TPCP map of the depolarizing noise on two qubits state $\rho'$ as follows:
\begin{equation}
    \mathcal{D}_2(\rho') = \left(1-\frac{15p}{16}\right) \mathcal{II}(\rho')  + \frac{p}{16} (\mathcal{IX}(\rho') + \cdots + \mathcal{ZZ}(\rho')). 
\end{equation}
The inverse map of $\mathcal{D}_2$ is written as follows \cite{guo2023Noise}:
\begin{equation}
\begin{split}
    \mathcal{D}_2^{-1}\left(\rho'\right) 
    =& \left(1 - \frac{15q}{16}\right) \mathcal{II}(\rho') \\
    &+ \frac{q}{16}  (\mathcal{IX}(\rho') + \mathcal{IY}(\rho') + \cdots + \mathcal{ZZ}(\rho')),
\end{split}
\end{equation}
where
\begin{equation}
    q = \frac{-p}{1-p}.  \label{eq:pec:2depo:q}
\end{equation}
The sampling cost is $\gamma=(1+7p/8)/(1-p)$, which gives that the insertion probabilities of Pauli gates (recovery gates) $IX, IY, \ldots$, and $ZZ$ are all the same value $p/(16+14p)$.

\section{Feed-Forward PEC} \label{sec:FFPEC}
We introduce Feed-Forward Probabilistic Error Cancellation (FFPEC) to obtain a more accurate expectation value by defining a new inverse map of noise taking a noise induced by recovery gate insertion.
The calculation process of FFPEC is the same as the PEC calculation process with the new inverse map instead of a conventional inverse map of noise.
In the following, we give the idea of FFPEC how to determine a new inverse map of noise with a well-known noise model, depolarizing.
Moreover, we show insertion probabilities of recovery gates and sampling costs for depolarizing for FFPEC.
Here, we assume that the noise caused by the gate operation and the probability $p$ of its occurrence are given and small $p \ll 0.5$.
Since the recovery gate added is also gate operation, the probability of noise induced by the recovery gate is required to be the same as $p$.

\begin{figure}[htbp]
    \centering
    \includegraphics[scale=0.5]{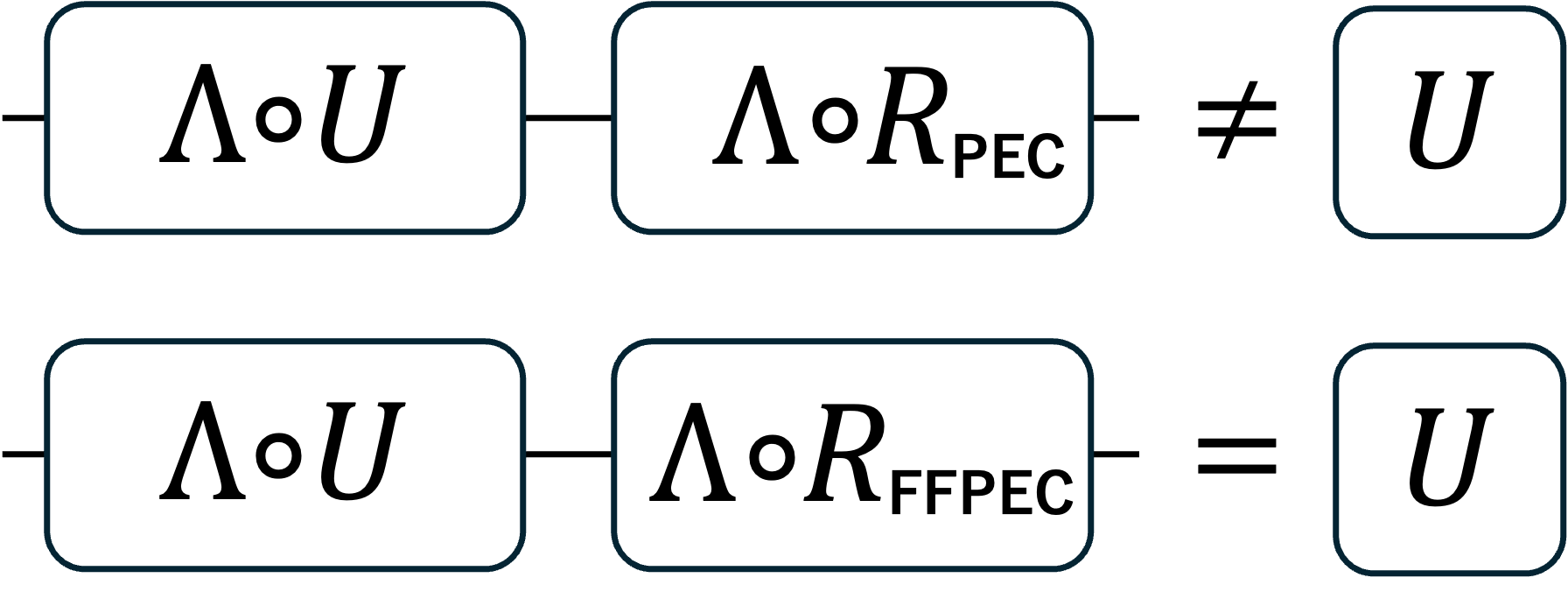}
    \caption{Conceptual diagram of FFPEC: a recovery gate $R_{\text{FFPEC}}$ of FFPEC can correct a noise $\Lambda$ even when the noise occurs after the recovery gate is performed unlike the standard PEC.
    }
    \label{fig:ffpec_concept}
\end{figure}

\subsection{Depolarizing for single qubit state}
We explain how to determine an inverse map of depolarizing on a single qubit state for FFPEC.
Let $\tilde{\mathcal{D}}_1^{-1}$ denote a noisy version of $\mathcal{D}_1^{-1}$ as follows:
\begin{equation}
     \mathcal{\tilde{D}}_{1}^{-1}(\rho) = \left(1- \frac{3q}{4}\right) \mathcal{I}(\rho)  + \frac{q}{4} (\tilde{\mathcal{X}}(\rho) + \tilde{\mathcal{Y}}(\rho) + \tilde{\mathcal{Z}}(\rho)),
\end{equation}
where $\tilde{\mathcal{X}}, \tilde{\mathcal{Y}}$, and $\tilde{\mathcal{Z}}$ are noisy version of $\mathcal{X}, \mathcal{Y}$, and $\mathcal{Z}$ for depolarizing, respectively.
These are represented as follows:
\begin{align}
    \tilde{\mathcal{X}}(\rho) &= \left(1- \frac{3p}{4}\right) \mathcal{X}(\rho)  + \frac{p}{4} (\mathcal{I}(\rho) +  \mathcal{Z}(\rho) + \mathcal{Y}(\rho)), \\
    \tilde{\mathcal{Y}}(\rho) &= \left(1- \frac{3p}{4}\right) \mathcal{Y}(\rho)  + \frac{p}{4} (\mathcal{Z}(\rho) + \mathcal{I}(\rho) + \mathcal{X}(\rho)), \\
    \tilde{\mathcal{Z}}(\rho) &= \left(1- \frac{3p}{4}\right) \mathcal{Z}(\rho)  + \frac{p}{4} (\mathcal{Y}(\rho) + \mathcal{X}(\rho) + \mathcal{I}(\rho)).
\end{align}
To obtain the parameter $q$ for FFPEC, equation $\mathcal{\tilde{D}}_{1}^{-1}(\mathcal{D}_1(\rho)) = \rho$ is solved.
Then we have
\begin{equation}
    q = \frac{-4p}{4-5p+p^2} = \frac{-4p}{(1-p)(4-p)}.  \label{eq:ffpec:1depo:q}
\end{equation}
The sampling cost is $\gamma = (4+p+p^2)/[(1-p)(4-p)]$ and the insertion probabilities of pauli gates $X$, $Y$, and $Z$ are all the same value $4p/(4+p+p^2)$.

\subsection{Depolarizing for two qubits state}
A modified inverse of depolarizing for two qubits is also determined by a similar way as for a single qubit state explained in the previous section.
Let $\tilde{\mathcal{D}}_2^{-1}$ denote a noisy version of $\mathcal{D}_2^{-1}$ as follows:
\begin{equation}
\begin{split}
     \mathcal{\tilde{D}}_{2}^{-1}(\rho') 
     =& \left(1-\frac{15q}{16}\right) \mathcal{II}(\rho')  \\
      &+ \frac{q}{16} (\widetilde{\mathcal{IX}}(\rho') + \widetilde{\mathcal{IY}}(\rho') + \cdots + \widetilde{\mathcal{ZZ}}(\rho')), 
\end{split}
\end{equation}
where $\widetilde{\mathcal{IX}}, \widetilde{\mathcal{IY}}, \ldots, \widetilde{\mathcal{ZZ}}$ are noisy versions of Pauli maps corresponding to $\mathcal{IX}, \mathcal{IY}, \ldots, \mathcal{ZZ}$.
Assuming that each $\widetilde{\mathcal{IX}}, \widetilde{\mathcal{IY}}, \ldots,\widetilde{\mathcal{ZZ}}$ is exposed to two qubits depolarizing noise, they can be represented as follows:

\begin{align}
    \widetilde{\mathcal{IX}}(\rho') &= \left(1-\frac{15p}{16}\right) \mathcal{IX}(\rho') + \frac{p}{16}\smashoperator[r]{\sum_{\mathcal{P} \in \{ \mathcal{I},\mathcal{X},\mathcal{Y},\mathcal{Z}\}^{\otimes 2} \backslash {\mathcal{\{IX\}}} }} \mathcal{P}(\rho'),\\
    \widetilde{\mathcal{IY}}(\rho') &= \left(1-\frac{15p}{16}\right) \mathcal{IY}(\rho') + \frac{p}{16}\smashoperator[r]{\sum_{\mathcal{P} \in \{ \mathcal{I},\mathcal{X},\mathcal{Y},\mathcal{Z}\}^{\otimes 2} \backslash {\mathcal{\{IY\}}} }} \mathcal{P}(\rho'),\\
    \vdots && \nonumber\\
    \widetilde{\mathcal{ZZ}}(\rho') &= \left(1-\frac{15p}{16}\right) \mathcal{ZZ}(\rho') + \frac{p}{16}\smashoperator[r]{\sum_{\mathcal{P} \in \{ \mathcal{I},\mathcal{X},\mathcal{Y},\mathcal{Z}\}^{\otimes 2} \backslash {\mathcal{\{ZZ\}}} }} \mathcal{P}(\rho').
\end{align}
To obtain the parameter $q$, equation $\tilde{\mathcal{D}}_2^{-1}(\mathcal{D}_2(\rho')) = \rho'$ is solved.
Then we have
\begin{equation}
    q = \frac{-16p}{16 -17p + p^2} = \frac{-16p}{(1-p)(16-p)}.  \label{eq:ffpec:2depo:q}
\end{equation}
The sampling cost is $\gamma = (16 + 13p + p^2)/[(1-p)(16-p)]$ and the insertion probabilities of Pauli gates $IX, \ldots$, and $ZZ$ are all the same value $p/(16+13p+p^2)$.
Note that the identity $I$ included in $IX, IY, \ldots,$ and $ZI$ gates is inserted as a recovery gate in this model as well as the standard PEC.

\section{Numerical Experiments} \label{sec:numexp}
We show some numerical experiments to investigate the accuracy of expectation values obtained by FFPEC.
To run noisy simulations, we use Qiskit (version 1.1.1)\cite{Qiskit}.

\subsection{Problem Settings}
\begin{figure*}[htbp]
    \begin{tabular}{ccc}
        \begin{minipage}{.32\hsize}
            \centering
            \includegraphics[width=0.85\columnwidth]{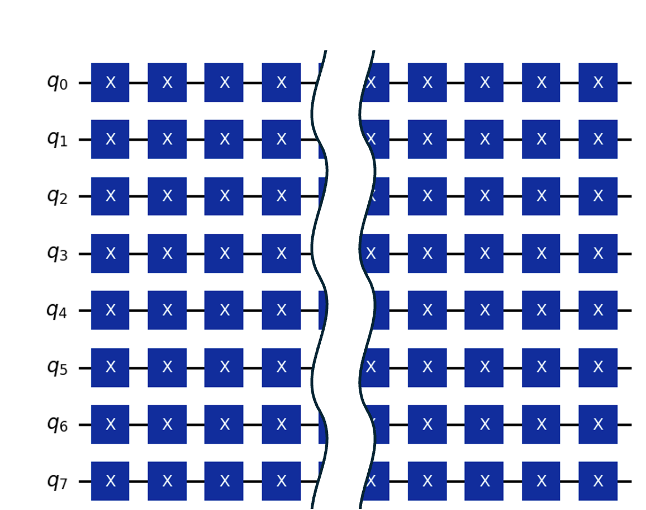}
            \subcaption{}
            \label{fig:circuit_a}
        \end{minipage}
        \begin{minipage}{.32\hsize}
            \centering
            \includegraphics[width=0.85\columnwidth]{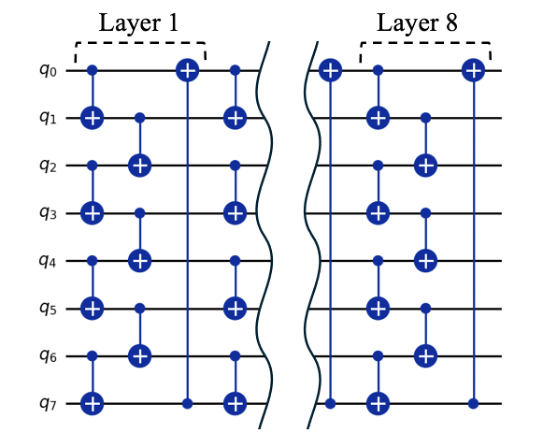}
            \subcaption{}
            \label{fig:circuit_b}
        \end{minipage}
        \begin{minipage}{.32\hsize}
            \centering
            \includegraphics[width=1.0\columnwidth]{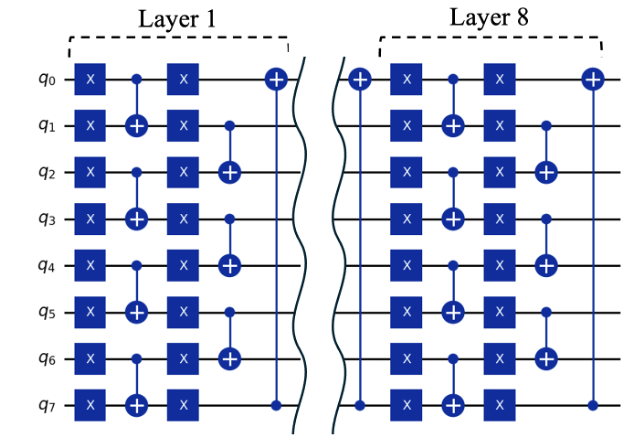}
            \subcaption{}
            \label{fig:circuit_c}
        \end{minipage}
    \end{tabular}
    \caption{Three types of 8 qubits quantum circuits used in experiments. The circuit \ref{fig:circuit_a} consists of 1600 $X$ gates. The circuit \ref{fig:circuit_b} consists of 8 layers, where one layer has 8 CNOT gates. The circuit \ref{fig:circuit_c} consists of 8 layers, where one layer has 16 $X$ gates and 8 CNOT gates.}
    \label{fig:circuits}
\end{figure*}

The depolarizing noise model is used with error rates $\{0.001, 0.0015, 0.002\}$ for single qubit gates and $\{0.01, 0.015, 0.02\}$ for two qubits gates, which are close to the error rates of current quantum computers.
The insertion probabilities of recovery gates for each error rate are summarized in Table \ref{table:adding_prob}.
For each error rate, the insertion probabilities in the FFPEC are slightly larger than those in the PEC, about $0.025\%, 0.038\%$ and $ 0.05\%$ for a single qubit gate, and $0.061\%, 0.091\%$ and $0.12\%$ for a two qubits gate.
To investigate these effects of the insertion probabilities on the accuracy of expectation values, experiments listed below are addressed using three types of quantum circuits consisting of basic gates as shown in Fig.~\ref{fig:circuits}:
\begin{itemize}
    \item[(a)] investigation of the accuracy of expectation values with error rates $\{0.001, 0.0015, 0.002\}$ for the $X$ gate with circuit [\ref{fig:circuit_a}],
    \item[(b)] investigation of the accuracy of expectation values with error rates $\{0.01, 0.015, 0.02\}$ for the CNOT gate with circuit [\ref{fig:circuit_b}] and,
    \item[(c)] investigation of the accuracy of expectation values using a circuit consisting of $X$ gates and CNOT gates [\ref{fig:circuit_c}] with error rates $\{ (0.001, 0.01), (0.0015, 0.015), (0.002, 0.02) \}$ for the $X$ gate and the CNOT gate.
\end{itemize}

\begin{table}[htbp]
\caption{Insertion probabilities of recovery gates for one noisy gate for each error rate.}
\label{table:adding_prob}
\begin{center}
\begin{tabular}{c|c|c|c|c}
Type                                 & Error Rate & PEC        & FFPEC    & \linestack{Relative \\difference} \\ \hline \hline
\multirow{3}{*}{Single qubit gate}   & 0.0010     & 0.0007496  & 0.0007498 & 0.0002497 \\ \cline{2-5} 
                                     & 0.0015     & 0.0011242  & 0.0011246 & 0.0003743 \\ \cline{2-5} 
                                     & 0.0020     & 0.0014985  & 0.0014993 & 0.0004988 \\ \hline \hline
\multirow{3}{*}{Two qubits gate}     & 0.0100     & 0.0092937  & 0.0092994 & 0.0006138 \\ \cline{2-5} 
                                     & 0.0150     & 0.0138803  & 0.0138930 & 0.0009123 \\ \cline{2-5} 
                                     & 0.0200     & 0.0184275  & 0.0184497 & 0.0012054 \\ \hline 
\end{tabular}
\end{center}
\end{table}

In the experiments, initial state $\ket{0}^{\otimes 8}$ and the observable $Z^{\otimes 8}$ are used.
The exact expectation values of the observable for each circuit are all $1$ to verify the influence of noise for expectation values obtained using the PEC and FFPEC.
$10^3$ of expectation values are sampled with $10^6$ shots to evaluate the accuracies (averages and standard deviations) of sampled expectation values using the PEC and FFPEC.

To more precisely evaluate the performance of the PEC and FFPEC, their analytical expectation values are used for each circuit in Fig.~\ref{fig:circuits}.
The analytical expectation values are obtained using the following TCPC map, which is included the noise $\Lambda$ to the recovery gate $\mathcal{R}$ based on Eq.~\eqref{eq:sum_Uk}:
\begin{equation}
    \prod_{k=1}^{N_G} \gamma^{(k)} \sum_{i_1, \cdots, i_{N_G}} \prod_{k=1}^{N_G} {\rm sgn}(q_{i_{k}}) \prod_{k=1}^{N_G} \sigma_{i_{k}} \prod_{k=1}^{N_G} \Lambda_k\mathcal{R}_{i_{k}} \Lambda_k \mathcal{U}_k.
\end{equation}

Table \ref{table:cost} shows the sampling overhead of the PEC and FFPEC for each error rate with respect to single qubit gates, two qubits gates, and circuits \ref{fig:circuit_a}, \ref{fig:circuit_b}, and \ref{fig:circuit_c}. 
The sampling overhead corresponds to the standard deviation, where both PEC and FFPEC have almost the same standard deviation.

\begin{table}[htbp]
\caption{Sampling overheads with each error rate.}
\label{table:cost}
\begin{center}
\begin{tabular}{c|c|c|c}
Type                                               & Error Rate & PEC      & FFPEC \\ \hline \hline

\multirow{3}{*}{$\gamma$ for single qubit gate}    & 0.0010 & 1.0015015 & 1.0015019 \\ \cline{2-4} 
                                                   & 0.0015 & 1.0022534 & 1.0022542 \\ \cline{2-4}
                                                   & 0.0020 & 1.0030060 & 1.0030075      

\\ \hline \hline
\multirow{3}{*}{$\gamma$ for two qubits gate}      & 0.0100 & 1.0189394 & 1.0189512 \\ \cline{2-4} 
                                                   & 0.0150 & 1.0285533 & 1.0285801 \\ \cline{2-4}
                                                   & 0.0200 & 1.0382653 & 1.0383132         

\\ \hline \hline
\multirow{3}{*}{$\gamma^{\text{tot}}$ of (a)}      & 0.0010 & 11.029799 & 11.036417 \\ \cline{2-4} 
                                                   & 0.0015 & 36.647750 & 36.697239 \\ \cline{2-4}
                                                   & 0.0020 & 121.80298 & 122.09551  

\\ \hline \hline
\multirow{3}{*}{$\gamma^{\text{tot}}$ of (b)}      & 0.0100 & 3.3227265 & 3.3251994 \\ \cline{2-4} 
                                                   & 0.0150 & 6.0605956 & 6.0707082 \\ \cline{2-4}
                                                   & 0.0200 & 11.059460 & 11.092156       

\\ \hline \hline
\multirow{4}{*}{$\gamma^{\text{tot}}$ of (c)}      & \linestack{single 0.0010 \\ two 0.0100} & 4.0262432 & 4.02943305 \\ \cline{2-4} 
                                                   & \linestack{single 0.0015 \\ two 0.0150} & 8.0842381 & 8.09860145 \\ \cline{2-4}
                                                   & \linestack{single 0.0020 \\ two 0.0200} & 16.240019 & 16.2911568 \\ \hline
\end{tabular}
\end{center}
\end{table}

\begin{table}[htbp]
\caption{Expectation values under the depolarising noise of circuits \ref{fig:circuit_a}, \ref{fig:circuit_b}, and \ref{fig:circuit_c} without the PEC and FFPEC}
\label{table:resluts_noisy}
\begin{center}
\begin{tabular}{c|c|c}
Type                                               & Error Rate & Expectation value \\ \hline \hline
\multirow{3}{*}{(a)}                               & 0.0010     & 0.2017 \\ \cline{2-3} 
                                                   & 0.0015     & 0.0906 \\ \cline{2-3}
                                                   & 0.0020     & 0.0406 

\\ \hline \hline
\multirow{3}{*}{(b)}                               & 0.0100     & 0.5256 \\ \cline{2-3} 
                                                   & 0.0150     & 0.3801  \\ \cline{2-3}
                                                   & 0.0200     & 0.2745       

\\ \hline \hline
\multirow{4}{*}{(c)}                               & \linestack{single 0.0010 \\ two 0.0100} & 0.4832 \\ \cline{2-3} 
                                                   & \linestack{single 0.0015 \\ two 0.0150} & 0.3351 \\ \cline{2-3}
                                                   & \linestack{single 0.0020 \\ two 0.0200} & 0.2320 \\ \hline
\end{tabular}
\end{center}
\end{table}

\subsection{Experimental results and  Discussions}

\begin{figure}[htbp]
    \centering
    \includegraphics[width=0.95\columnwidth]{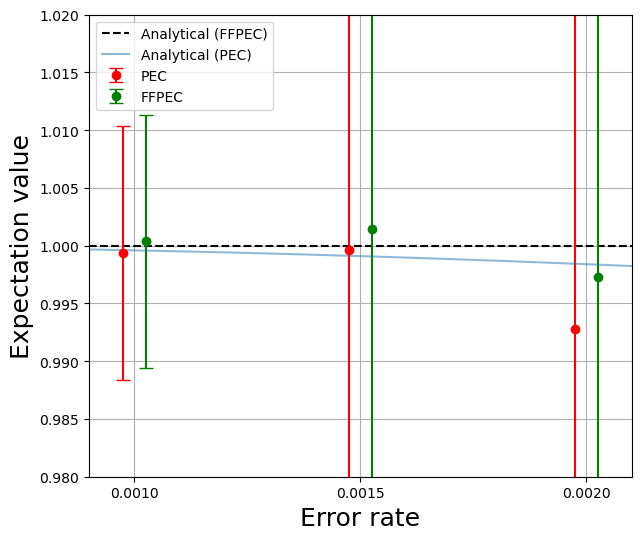}
    \caption{Results with circuit \ref{fig:circuit_a}. The standard deviations are $0.03680$ at an error rate of $0.0015$ and $0.1231$ at an error rate of $0.002$.}
    \label{fig:result_a}
\end{figure}

\begin{figure}[htbp]
    \centering
    \includegraphics[width=0.95\columnwidth]{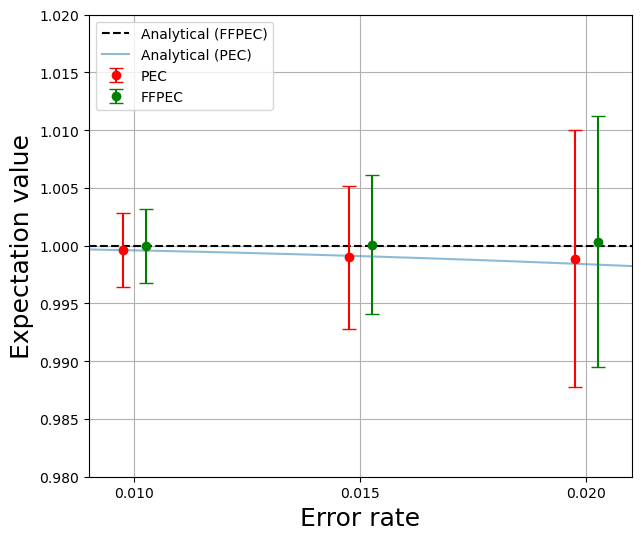}
    \caption{Results with circuit \ref{fig:circuit_b}.}
    \label{fig:result_b}
\end{figure}

\begin{figure}[htbp]
    \centering
    \includegraphics[width=0.95\columnwidth]{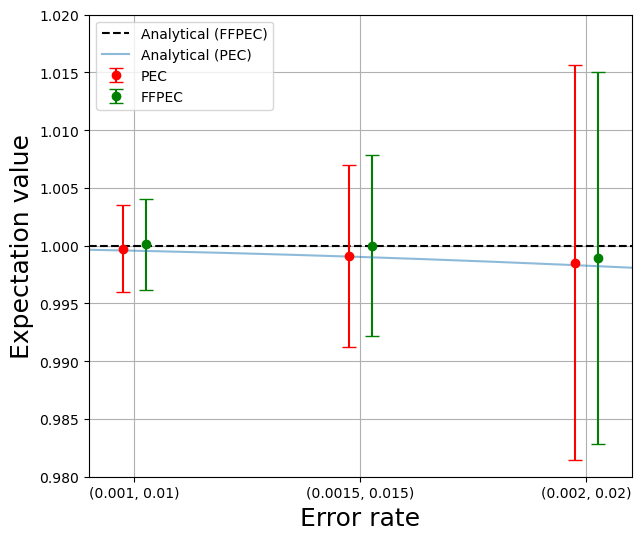}
    \caption{Results with circuit \ref{fig:circuit_c}.}
    \label{fig:result_c}
\end{figure}

Figures~\ref{fig:result_a}, \ref{fig:result_b}, and \ref{fig:result_c} show mitigation results for circuits \ref{fig:circuit_a}, \ref{fig:circuit_b}, and \ref{fig:circuit_c}, respectively.
The horizontal axis represents error rates, and the vertical axis represents expectation values obtained using the PEC and FFPEC.
The red lines show the numerical results of PEC and the green lines show those of FFPEC by samplings.
The blue line represents analytical expectation values obtained by PEC, while the black dashed line represents those of FFPEC.
Although the analytical expectation values of PEC do not match the exact expectation value of $1$ due to the influence of noise by inserting the recovery gate, the analytical expectation value of FFPEC matches the exact one.
Here, the scale on the vertical axis of all results in three figures is consistent to allow comparison of analytical expectation values with those of numerical ones.
In Fig.~\ref{fig:result_a}, the standard deviations are $0.03680$ at error rate $0.0015$ and $0.1231$ at error rate $0.002$.
In all results, the standard deviations of the PEC and FFPEC are almost the same because they have the same sampling overhead summarized in Table~\ref{table:cost}.
In addition, both the PEC and FFPEC can provide improved expectation values as shown in Figs.~\ref{fig:result_a}, \ref{fig:result_b}, and \ref{fig:result_c}, where expectation values without mitigation are summarized in Table~\ref{table:resluts_noisy}.

The results from the PEC and FFPEC shown in Fig.~\ref{fig:result_a} show that, as the error rate increases, the standard deviations increase and the averages shift away from the exact expectation value 1.
This is because $10^6$ shots are not enough for convergence of expectation values of the circuit \ref{fig:circuit_a} with large sampling overheads.
For the error rate $0.001$, which is approximately that of current quantum computers of single qubit gate, both the PEC and FFPEC have the almost same absolute error from the exact value of $1$, which is $9.256 \times 10^{-4}$.
This observation implies that the performance of FFPEC is the same as that of the PEC.
If more than $10^6$ shots are used, FFPEC can provide a more accurate expectation value than the PEC.

In the results shown in Fig.~\ref{fig:result_b}, the average values obtained using both the PEC and FFPEC converge to their analytical expectation values, respectively.
Focusing on the error rate $0.01$, which is approximately that of current quantum computers of two qubits gate (e.g. CNOT gate), the absolute error between the exact expectation value and the average one is $3.858 \times 10^{-4}$ for the PEC and $5.847 \times 10^{-5}$ for FFPEC.
Thus, FFPEC can provide a more accurate expectation value than the PEC when circuits like Fig.~\ref{fig:circuit_b} are used.

In the results shown in Fig.~\ref{fig:result_c}, the average values obtained using PEC converge to their analytical expectation values.
The average values of FFPEC also converge to the analytical expectation values, except for the result at the error rate $(0.002, 0.02)$.
The reason why the result at the error rate does not converge to the analytical expectation value is the shortage of shots.
Focusing on the error rate $(0.001, 0.01)$, which is approximately that of current quantum computers of single and two qubits gate, the absolute error between the exact expectation value and the average value is $2.743 \times 10^{-4}$ for the PEC and $1.084 \times 10^{-4}$ for FFPEC.
Thus, FFPEC can provide a slightly more accurate expectation value than the PEC.

Since practical quantum circuits consist of single qubit gates and CNOT gates, as shown in the circuit \ref{fig:circuit_c}, the results of the error rate $(0.001, 0.01)$ in Fig.~\ref{fig:result_c} are expected results on current quantum computers, although thermal relaxation errors and readout errors exist.
FFPEC can provide slightly more accurate expectation values than the PEC as stated above.
The error mitigation capability of the PEC is limited due to the noise induced by inserting recovery gates, but FFPEC has no such limitation.
Thus, the more shots are taken to obtain a highly accurate expectation value, FFPEC can provide a more accurate one than the PEC.
For the practical implementation of FFPEC, there are challenges:
(i) to address other errors such as thermal relaxation errors and readout errors,
(ii) to improve error inverse maps based on the working of gates on real hardware such as the identity gate is not executed, a two qubits gate is executed as per single qubit gate if the two qubits gate can be decomposed to its tensor product, the $Z$ gate can be performed as virtual \cite{PhysRevA.96.022330}, and
(iii) to reduce the number of shots to provide expectation values with the required accuracy.
In addition, both the PEC and FFPEC can sometimes yield expectation values that exceed the range $[-1, 1]$, as shown in the numerical experiments
For algorithms utilizing the PEC and FFPEC, it would be unsuitable to use values outside this range.
Accordingly, it would be better to correct to $1$ or $-1$ when such values are obtained.

\section{Conclusion} \label{sec:conculusions}
We explore an improved version of the PEC to achieve zero bias for the estimator of expectation values even with recovery gates inducing noises that is a realistic situation.
We propose Feed-Forward Probabilistic Error Cancellation (FFPEC) method, which can provide a zero bias estimator even in such a situation.
The key idea of FFPEC is to define a new inverse map of noise to cancel out noises induced by recovery gates.
This idea would be a simple way for quantum error mitigation techniques such as the PEC, which utilize the insertion of recovery gates.

To demonstrate the idea of FFPEC, new versions of inverse maps for the depolarizing noise on single qubit and two qubits gates are provided, and analytical evaluation is carried out.
Moreover, numerical experiments are also carried out and compared with analytical results.
Experimental results show that FFPEC can provide more accurate expectation values than the PEC on depolarizing noise.
When the error rate is small, the performance of FFPEC is the same as the PEC.
In situations where a high accuracy is required, FFPEC can be a promising tool.
However, in practice, it is important to construct FFPEC considering not only depolarizing noises but also other noises and actual operations.
We focus on constructing practical inverse maps that take these factors into account.

\bibliographystyle{IEEEtran}
\bibliography{main}

\end{document}